\documentclass[9pt]{article}
\usepackage{spconf,amsmath,graphicx}

\usepackage{float}

\title{MusicMamba: A Dual-Feature Modeling Approach for Generating Chinese Traditional Music with Modal Precision}
\name{
Jiatao Chen$^*$ \hspace{1em} Xing Tang$^*$ \hspace{1em} Tianming Xie$^*$ \hspace{1em}  Jing Wang$^\dagger$\hspace{1em} Wenjing Dong$^*$ \hspace{1em} Bing Shi$^*$ \vspace{-1.25ex}}
\address{$^*$School of Computer Science and Artificial Intelligence, Wuhan University of Technology, Wuhan, China\\
$^\dagger$School of Computer Science, Hubei University of Technology, Wuhan, China\\
\thanks{Jiatao Chen and Tianming Xie are co-first authors, and Xing Tang is the corresponding author. This work was supported in
part by the National Natural Science Foundation of China under Grant
62272356, Grant 62302155, and Grant 62372344 and in part by the Natural
Science Foundation of Chongqing under Grant CSTB2022NSCQ-MSX1414.}}

\usepackage[T1]{fontenc} 
\usepackage[utf8]{inputenc} 
\usepackage[hidelinks,pagebackref,colorlinks=true,backref=false]{hyperref}
\usepackage{url}
\usepackage{color}

\usepackage{cite}
\usepackage{inconsolata}
\usepackage{amsmath}
\usepackage{amssymb}
\usepackage{booktabs}
\usepackage{microtype}
\usepackage{enumitem}
\usepackage[noabbrev,capitalise]{cleveref}
\usepackage{bm}
\usepackage{bbm}
\usepackage{multirow}
\usepackage{tabularx}

\usepackage{cleveref}
\crefname{figure}{fig.}{figures}
\Crefname{figure}{Fig.}{Figures}

\setlist{parsep=0ex,topsep=0.5ex,itemsep=0ex,leftmargin=2em}
\crefformat{footnote}{#2\footnotemark[#1]#3}
\crefrangeformat{footnote}{#3\footnotemark[#1]#4--#5\footnotemark[#2]#6}
\crefmultiformat{footnote}{#2\footnotemark[#1]#3}{\textsuperscript{,}#2\footnotemark[#1]#3}{\textsuperscript{,}#2\footnotemark[#1]#3}{\textsuperscript{,}#2\footnotemark[#1]#3}
\graphicspath{{figs/}}

\newcolumntype{C}{>{\centering\arraybackslash}X}
\definecolor{darkred}{rgb}{0.6,0.0,0.0}
\definecolor{darkblue}{rgb}{0.0,0.0,0.75}
\definecolor{myblue}{rgb}{0,0.3,0.6}
\hypersetup{linkcolor=myblue,urlcolor=myblue,citecolor=myblue,anchorcolor=myblue}

\begin{document}
\ninept
\maketitle

\begin{abstract}
In recent years, deep learning has advanced the MIDI domain, solidifying music generation as a key application of artificial intelligence. However, most research focuses on Western music, facing challenges in generating Chinese traditional melodies, particularly in capturing modal characteristics and emotional expression. To address this, we propose the Dual-Feature Modeling Module, which integrates the long-range modeling of the Mamba Block with the global structure capturing of the Transformer Block. Additionally, we introduce the Bidirectional Mamba Fusion Layer, which integrates local details and global structures through bidirectional scanning, enhancing sequence modeling. Building on this, we propose the REMI-M representation to better capture and generate modal information in melodies. To support this, we developed FolkDB, a high-quality Chinese traditional music dataset covering over 11 hours of music. Experimental results show our architecture excels in generating melodies with Chinese traditional music characteristics, offering a new solution for music generation.

\end{abstract}

\setlength{\abovedisplayskip}{1ex}
\setlength{\belowdisplayskip}{1ex}
\setlength{\abovedisplayshortskip}{1ex}
\setlength{\belowdisplayshortskip}{1ex}

\section{Introduction}
\label{sec:introduction}
Recent advances in deep learning have significantly impacted the MIDI domain, making music generation a key application of artificial intelligence. Melody generation, a central task in music composition, involves creating musical fragments through computational models and presents more challenges than harmony generation and arrangement. A successful model must capture essential features like pitch and rhythm while producing melodies that align with specific styles and emotions. However, most existing methods, whether based on Recurrent Neural Networks \cite{Hakimi2020,Roberts2018,Johnson2017,Wang} or Transformer architectures \cite{Huang2018, Madaghiele2021, Ren2020, Zhang2023a}, struggle with the complexity and structure of melodies. For example, while \cite{Zixun2021} generated long-term structured melodies, Transformers excel in capturing global dependencies, demonstrating strong performance across various melody tasks \cite{Wu2024}.

Several studies have integrated music theory into the generation process. For instance, \cite{Zhu2018} introduced chord progressions for melody generation, \cite{Wang2020} controlled polyphonic music features through chords and textures, and \cite{Huang2020} improved beat structure representation. Additionally, \cite{Madaghiele2021} generated harmonious jazz melodies by adjusting harmonic and rhythmic properties, while other works have explored structured music generation using note-to-bar relationships \cite{Ruette2023} and melody skeletons \cite{Zhang2023}.

Meanwhile, State Space Models (SSMs) have advanced in modeling long-sequence dependencies, particularly in capturing global musical structures. Models like S4 \cite{Gu} and S5 \cite{Smith2023} have significantly improved parallel scanning efficiency through new state space layers. Mamba \cite{Gu2023}, a successful SSM variant, enhances parallel computation and has been applied across various fields, including visual domains with VMamba \cite{Liu2024} and large-scale language modeling with Jamba \cite{Lieber2024}. Recognizing Mamba's potential in sequence modeling, we applied it to symbolic music generation.

\begin{figure}
    \centering
    \includegraphics[width=\linewidth]{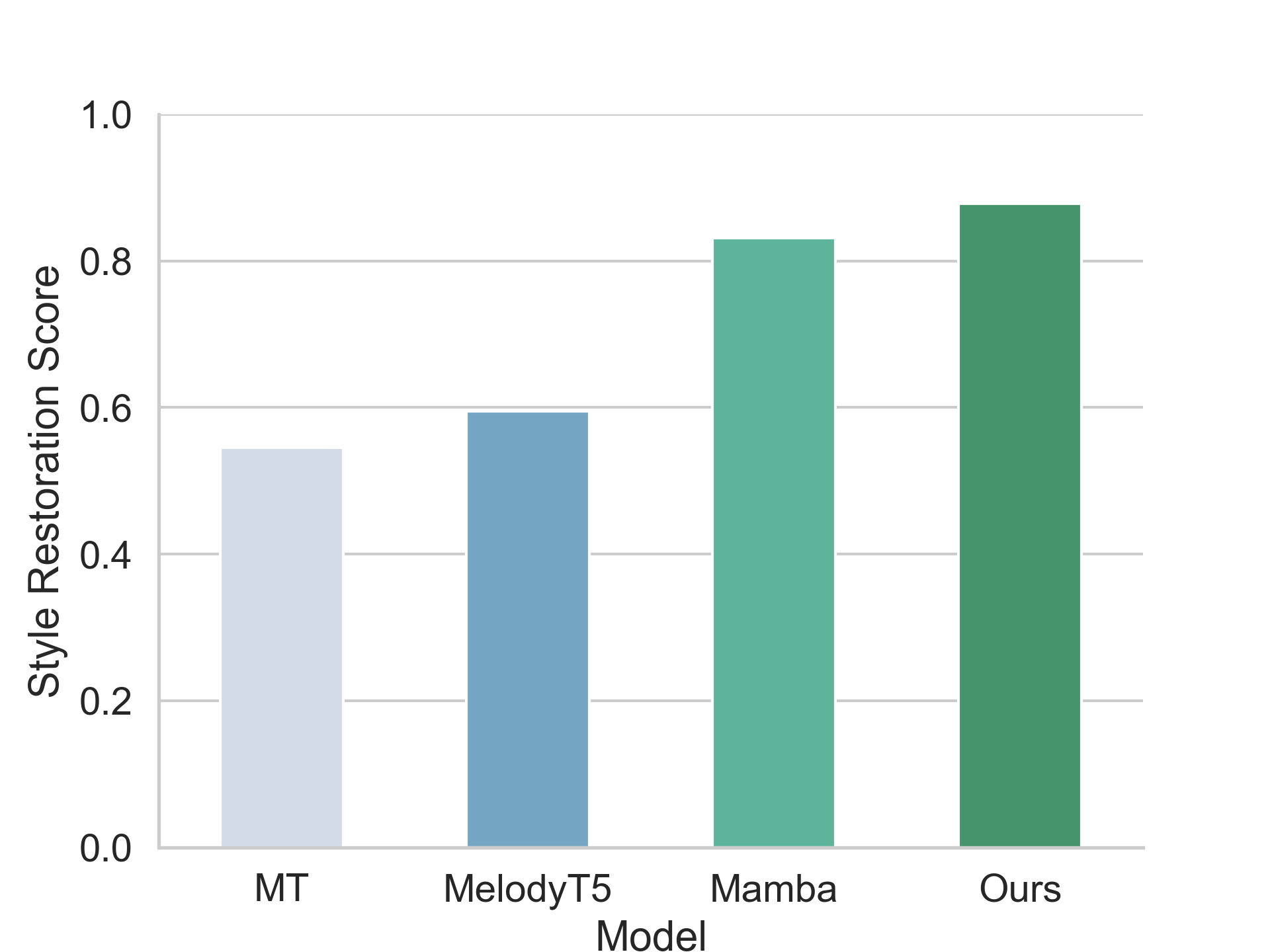}
    \caption{The scores of various models for replicating Chinese folk music with the specific style metric.}
    \label{fig:style_restoration_score}
\end{figure}

However, these methods primarily focus on Western music and struggle with generating Chinese traditional melodies. While they can produce smooth melodies, they often align with modern styles, failing to capture the unique contours and rhythms of Chinese traditional music. As shown in \Cref{fig:style_restoration_score}, existing methods underperform in preserving the stylistic elements of Chinese music. Modes play a central role in Chinese melodies, determining note selection and arrangement, while conveying specific emotions and styles \cite{Zhang2022}. Due to significant differences in scales, pitch relationships, and modal structures between Western and Chinese music, these methods fail to capture these modal characteristics, leading to discrepancies in style and emotional expression \cite{Hao2023}. The lack of high-quality Chinese traditional music datasets further limits their effectiveness.

\newpage
To address these issues, we propose a new architecture, the Dual-Feature Modeling Module, which combines the long-range dependency modeling of the Mamba Block with the global structure capturing of the Transformer Block. We also design the Bidirectional Mamba Fusion Layer, which integrates local details and global structures through bidirectional scanning, enhancing complex sequence modeling. This comprehensive architecture enables the generation of Chinese traditional music with complex structures and coherent melodies. Specifically, our contributions are:

\begin{itemize}
    \item \textbf{Mamba architecture to the MIDI domain.} We apply the Mamba architecture to MIDI music generation, proposing the Dual-Feature Modeling Module, which combines the strengths of Mamba and Transformer Blocks. Through the Bidirectional Mamba Fusion Layer, we integrate local details with global structures, achieving excellent performance in long-sequence generation tasks.
    
    \item \textbf{REMI-M Representation.} We extend the REMI representation with REMI-M, introducing mode-related events and note type indicators, allowing the model to more accurately capture and generate modal information in melodies.
    
    \item \textbf{FolkDB.} We create a high-quality Chinese traditional music dataset, FolkDB, designed for studying Chinese traditional music. With over 11 hours of music covering various styles, FolkDB fills a gap in existing datasets and provides a foundation for further research.
\end{itemize}

\section{Proposed Method}
\label{sec:method}

\subsection{Problem Formulation}
\label{sec:formulation}

In melody generation, the condition sequence is typically defined as $x_{1:t} = [x_1,...,x_t]$ and the target sequence as $y_{1:k}=[y_1,...,y_{k}]$, where $k>t$. The prediction of the $j$-th element in the target sequence can be expressed as $y_j|[x_1,...,x_t]\sim{p(y_j|x_1,...,x_t)}$ where $p(y_j|x_1,...,x_t)$ represents the conditional probability distribution of $y_j$ given the condition sequence.
Chinese traditional music often includes various modes. For example, in pentatonic modes, if a note $N_i\in\{C,D,E,G,A\}$ serves as the tonic note, then the following notes, if they follow a specific interval relationship, form a mode $M$. Therefore, to generate Chinese music with mode characteristics, the target sequence can consist of multiple modes and transition notes, represented as $y_{1}=(M_1,f(M_1),M_2,f(M_2),...,M_l,f(M_l))$, where $M_i$ corresponds to a subsequence of notes within a specific mode. where $M_i$ corresponds to a subsequence of notes within a specific mode, and $M_i$ represents the transition note sequence following $M_i$. The task of generating melodies with Chinese modes can ultimately be formulated as the following autoregressive problem:
\begin{equation}
    p(\mathbf{y}|\mathbf{x}, M) = \prod_{i=1}^{l} p(M_i|C_i) \cdot p(f(M_i)|C'_i)\,,
\end{equation}
where $M$ is the collection of multiple modes, $ C_i = (\mathbf{x}, \mathbf{y}_{<i})$ and $C'_i = (\mathbf{x}, \mathbf{y}_{\leq i})$. During the step-by-step generation of notes, the corresponding mode sequence $M_i$ is generated first, followed by the generation of the transition note sequence $f(M_i)$ based on the mode sequence.

\subsection{REMI-M Representation}
\label{sec:representation}

In generating traditional Chinese music, mode generation is a crucial and complex component. Chinese music often features intricate modal structures, such as pentatonic and heptatonic scales, where the selection and transition of modes are vital to the style and expression of the music. However, existing music representation methods face significant limitations in capturing and generating these modal structures. Although the REMI representation \cite{Huang2020} effectively captures rhythm, pitch, and velocity information through events such as bar, position, tempo, and note, it struggles with complex modal structures, particularly when handling the dynamic modes in Chinese music.

To address this issue, we extend REMI by introducing two new events in the REMI-M representation to explicitly describe modes:

\begin{itemize}
    \item \textbf{Note type event.} Distinguishes between mode notes and transition notes, helping the model to more accurately capture modal information.

    \item \textbf{Mode-related events.} Include the start, end, and type of mode, enabling REMI-M to explicitly annotate and generate modal changes in the music.
\end{itemize}

\begin{figure}
    \small
    \centering
    \includegraphics[width=.99\linewidth]{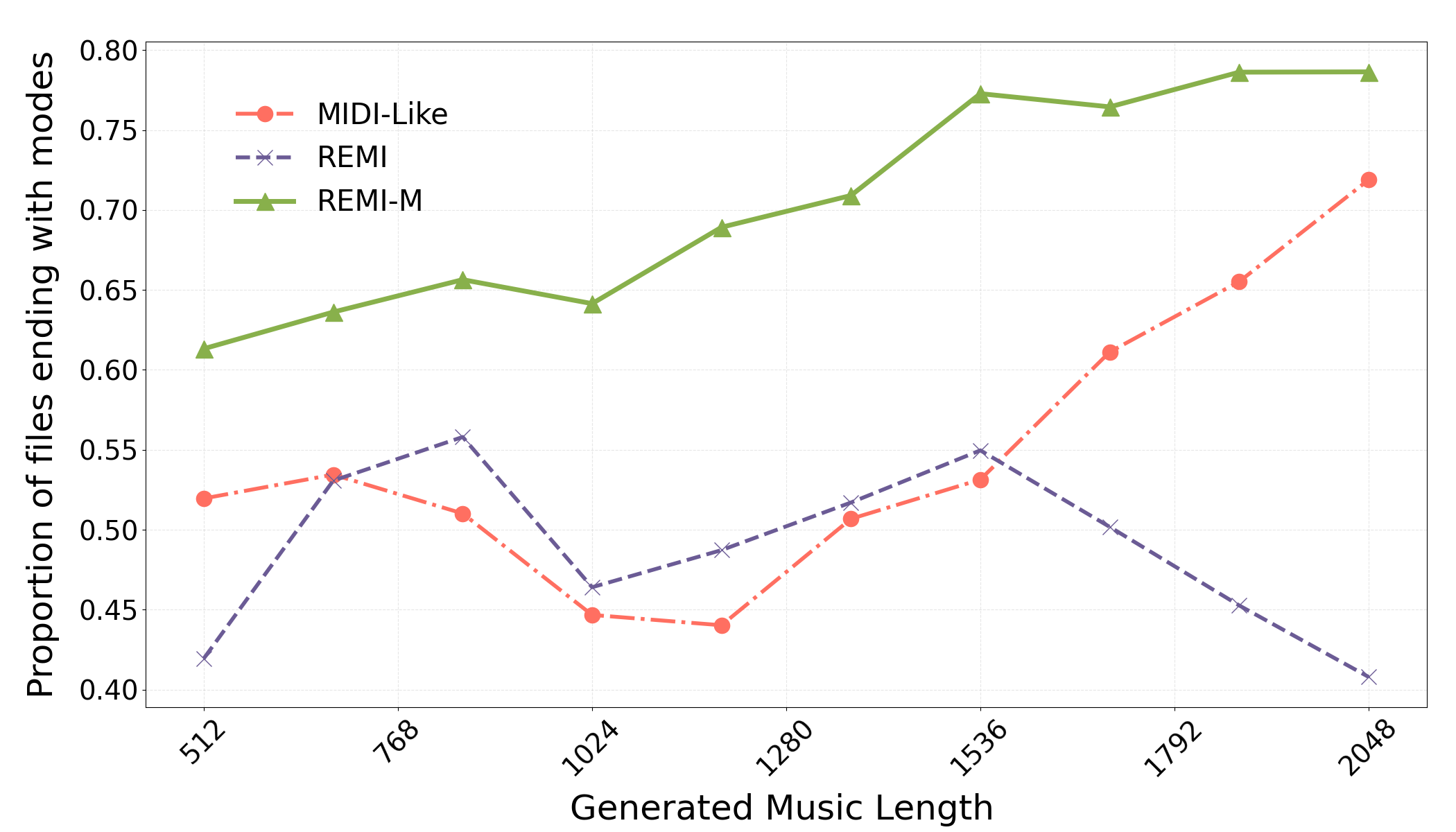}
    \caption{The proportion of generated music sequences with modes using three encoding schemes: MIDI-Like, REMI, and REMI-M. }
    \label{fig:representation_compare}
\end{figure}

As shown in \Cref{fig:representation_compare}, the original REMI and MIDI-Like encodings result in low mode generation rates, whereas REMI-M demonstrates significant improvements, achieving mode generation rates exceeding $0.8$ across all tested music lengths. These enhancements allow REMI-M to better handle complex modal structures, significantly improving the stylistic consistency and theoretical accuracy in generated music.

\subsection{Model}
\label{sec:model}
\begin{figure}
    \small
    \centering
    \includegraphics[width=.83\linewidth]{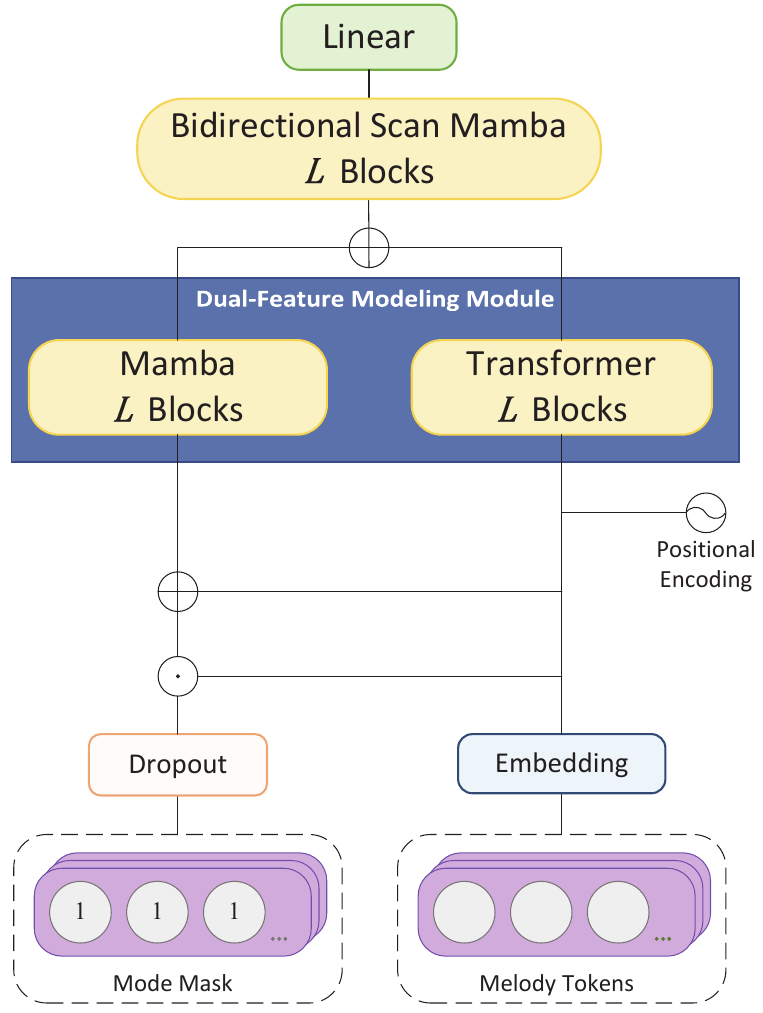}
    \caption{Illustration of the proposed MusicMamba model.}
    \label{fig:model}
\end{figure}

In music generation tasks, it is crucial to capture both local melodic details and global musical structure dependencies within long contexts. To achieve this, as shown in \Cref{fig:model}, we designed a hierarchical feature extraction and integration architecture named the Dual-Feature Modeling Module, which combines the long-range dependency modeling capability of the Mamba Block with the global structure capturing ability of the Transformer Block.

\textbf{Dual-Feature Modeling Module.} In music sequence generation, both melodic details (like note variations and modal transitions) and overall structure (such as phrases and repetition patterns) are crucial. Traditional architectures often struggle to capture these levels of features simultaneously. Let $\mathbf{H}$ represent the feature matrix. The Mamba Block captures melodic details and modal dependencies by computing a dot product between the mode mask and melody tokens, generating the feature representation $\mathbf{H}_1$. This provides essential long-range and local information for integration. The Transformer Block primarily models global structural information, processing input melody embeddings with positional encoding to obtain the structural representation $\mathbf{H}_2$.

\textbf{Bidirectional Mamba Fusion Layer.} To integrate the outputs of the Mamba Block and Transformer Block, we introduce the Bidirectional Mamba Fusion Layer. This layer simultaneously receives the long-range features $\mathbf{H}_1$ generated by the Mamba Block and the global features $\mathbf{H}_2$ generated by the Transformer Block. Through a bidirectional scanning mechanism, the forward and backward features are processed separately to obtain $F_{\text{forward}}$ and $F_{\text{backward}}$. Then, self-attention is applied to the forward and backward features to extract key information:

\begin{equation}
    F_1 = \text{Attention}(F_{\text{forward}}), \quad F_2 = \text{Attention}(F_{\text{backward}})\,.
\end{equation}
Next, these two directional features are concatenated to obtain the fused feature $\mathbf{H}_{\text{fusion}}$, and further processed by a linear layer:
\begin{equation}
    \text{Output}= \text{Linear}(\mathbf{H}_{\text{fusion}})\,.
\end{equation}
The fused feature $\mathbf{H}_{\text{fusion}}$ combines the long-range dependencies and global structure of the melody, providing complete information support for generating complex and coherent music sequences. Finally, the linear layer maps the fused features to the output space, generating the final music sequence.

\section{Experiments}
\label{sec:experiment}

\subsection{Implementation Details}
\subsubsection{Dataset}
We use two datasets: the POP909 dataset\cite{Wang*2020} and a self-collected Chinese Traditional Music dataset (referred to as the FolkDB). The POP909 dataset contains rich musical information, particularly in chords and melodies. Pre-training on this dataset allows the model to learn fundamental musical structures and elements, helping it to adapt more quickly to our FolkDB. Additionally, to address the lack of cultural diversity in the POP909 dataset, we have compiled a dataset of approximately 300 Chinese traditional music pieces. This dataset contains about 11 hours of piano MIDI works, featuring traditional modes such as the pentatonic and heptatonic scales, showcasing the diverse styles and modal characteristics of Chinese music.

In terms of data preprocessing, since the original data consists of single-track Chinese traditional music melodies, we perform additional processing on the self-collected Chinese traditional music dataset to ensure that the model can effectively capture and generate music with Chinese cultural characteristics. The specific steps are as follows:

\begin{itemize}
    \item \textbf{Tonic Track Extraction.} We employ the tonic extraction framework mentioned in the Wuyun model\cite{Zhang2023}. This framework uses a layered skeleton-guided approach, first constructing the skeleton of the melody and then extending it. 
    
    \item \textbf{Mode Detection and Annotation.} After extracting the tonic track, we conduct mode detection on the melody using interval relationships. By analyzing the intervals between each pair of tonic notes and leveraging the knowledge-enhanced logic within the Wuyun model, we obtained the mode track for each piece of music. 
\end{itemize}

\subsubsection{Model Settings}
We adopt an architecture based on the MambaBlock2 module\cite{Gu2023}, with the model's hidden dimension set to 256 and the feedforward network's intermediate layer dimension set to 1024. Additionally, we employ GatedMLP to enhance the model's nonlinear representation capabilities. During training, we use the Adam optimizer with an initial learning rate of $2 \times 10^{-4}$, dynamically adjusted via the LambdaLR scheduler. The training data is processed in batches, with each batch containing 8 samples and a fixed sequence length of 512 tokens. We use the cross-entropy loss function to measure the difference between the model's predictions and the target labels. This loss function is defined as follows:
\begin{equation}
    \text{Loss} = L_{\text{CE}} - \lambda_1L_{\text{NT}} - \lambda_2L_{\text{MR}}\,.
\end{equation}
Among these, \(L_{\text{CE}}\) is used to focus on the model's ability to accurately predict musical events, while \(L_{\text{NT}}\) and \(L_{\text{MR}}\) correspond to note type events and mode-related events, respectively. $\lambda_1$ and $\lambda_2$ are used to balance the contributions of the two losses, with both values typically ranging between 0 and 1.

\begin{table*}[ht]
    \footnotesize
    \vspace{-1.5ex}
    \caption{Performance comparison of our proposed model against the baseline models.}
    \label{tab:performance}
    \vspace{1ex}
    \centering
    \begin{tabular}{lccccccc}
        \toprule
        \textbf{Model} & \textbf{Average Groove} & \textbf{Average Style} & \textbf{Mode} & \multicolumn{3}{c}{\textbf{Subjective listening test results}} \\
        \cmidrule(lr){5-7}
        & \textbf{Consistency (\%)} & \textbf{Consistency (\%)} & \textbf{Consistency (\%)} & \textbf{Coherence} & \textbf{Richness} & \textbf{Style} \\
        \midrule
        MusicTransformer\cite{Huang2018} & 42.3 & 65.4 & 37.5 & 7.48 & 7.59 & 6.55 \\
        Mamba\cite{Gu2023} & 44.3 & 73.0 & 62.1 & 7.73 & 7.43 & 7.51 \\
        \cmidrule(lr){1-7}
        MusicTransformer\cite{Huang2018} & 45.2 & 69.7 & 53.6 & 7.06 & 7.49 & 7.67 \\
        MelodyT5\cite{Wu2024} & 51.8 & 67.9 & --- & 7.21 & 7.65 & 6.86 \\
        \textbf{Ours} & \textbf{59.9} & \textbf{85.3} & \textbf{66.1} & \textbf{7.91} & \textbf{7.72} & \textbf{8.26}  \\
        \bottomrule
    \end{tabular}
\end{table*}

\subsection{Objective Evaluation}
\subsubsection{Metric}
To evaluate our music generation model, we select the following four objective metrics: Pitch Class Entropy, Groove Consistency\cite{Wu2020, Mogren2016, Sturm2016}, Style Consistency, and Mode Consistency. 

\begin{itemize}
    \item \textbf{Pitch Class Entropy.} This metric reflects the diversity of pitch distribution, with higher entropy indicating a more dispersed distribution of generated notes, while lower entropy indicates a more concentrated distribution.

    \item \textbf{Groove Consistency.} Higher groove consistency indicates less variation in rhythm, resulting in a smoother, more stable musical flow.

    \item \textbf{Style Consistency.} A higher style consistency score indicates that the generated music aligns more closely with the expected style.

    \item \textbf{Mode Consistency.} This metric evaluates whether the notes in the generated music conform to the predefined mode structure. We improved the traditional scale consistency metric \cite{Wu2020, Mogren2016} to better align with the modal characteristics of Chinese folk music. The specific formula is as follows:
\end{itemize}


\begin{equation}
    \text{Consistency Score} = \frac{|\mathcal{P}_{\text{melody}} \cap \mathcal{P}_{\text{scale}}|}{|\mathcal{P}_{\text{melody}}|} \times 100\%\,,
\end{equation} 

Here, $\mathcal{P}_{\text{melody}}$ represents the set of melody notes, $\mathcal{P}_{\text{scale}}$ represents the set of scale notes, ${|\mathcal{P}_{\text{melody}} \cap \mathcal{P}_{\text{scale}}|}$ denotes the size of the intersection between the melody note set and the scale note set, and $\mathcal{P}_{\text{melody}}$ represents the size of the melody note set. The consistency score is determined by calculating the overlap ratio between the sets of notes in the melody and scale tracks.

\begin{table}[ht]
    \footnotesize
    \vspace{-1.5ex}
    \caption{Comparison of Average Pitch Entropy among models.}
    \label{tab:pitch-entropy}
    \vspace{1ex}
    \centering
    \begin{tabular}{lcc}
        \toprule
        \textbf{Model} & \textbf{Average Pitch Entropy} \\
        \midrule
        Ground truth & 3.831 \\
        \midrule
        MusicTransformer\cite{Huang2018} & 3.124 \\
        Mamba\cite{Gu2023} & 3.647 \\
        \cmidrule(lr){1-2}
        MusicTransformer\cite{Huang2018} & 3.404 \\
        MelodyT5\cite{Wu2024} & 3.530 \\
        \textbf{Ours} & \textbf{4.070} \\
        \bottomrule
    \end{tabular}
\end{table}

\subsubsection{Results}
\begin{figure}
    \small
    \centering
    \includegraphics[width=\linewidth]{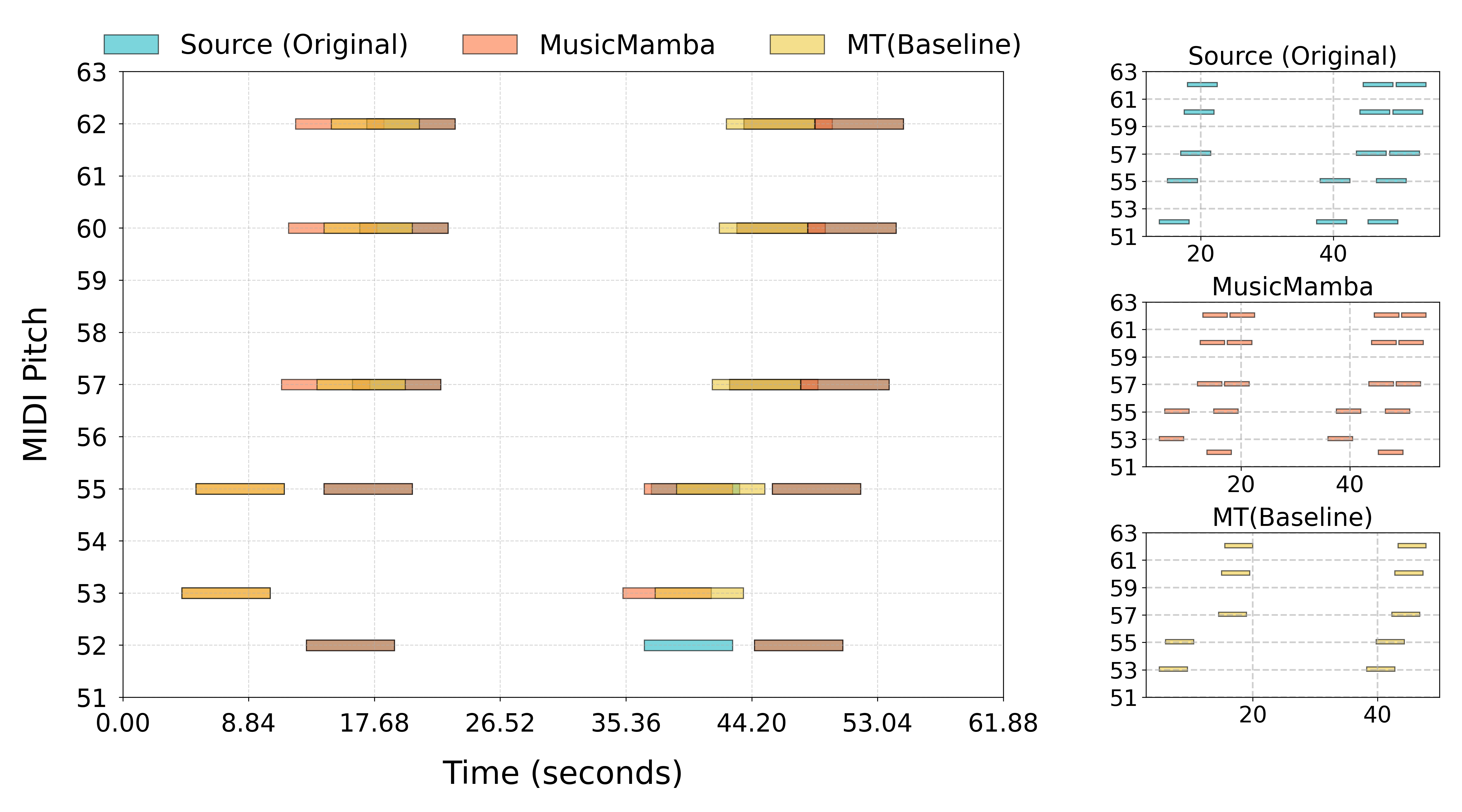}
    \caption{Mode distributions of the original MIDI sequence (Source) and sequences generated by MusicMamba and MT (MusicTransformer). The left panel illustrates the overlapping mode distributions, while the right panel presents the individual mode distributions for each sequence.}
    \label{fig:mode_generated}
\end{figure}

Before introducing the objective indicator test results, we test the key restoration of each model, and it is clear from the \Cref{fig:mode_generated} that MusicMamba not only effectively restores the key in the original sequence, but also introduces additional key changes, while MusicTransformer, although it captures some keys, is not as comprehensive and diverse as MusicMamba. Experimental results show that MusicMamba is better at generating melodies with traditional Chinese music styles, and can generate richer and more consistent sequences.

We conduct two sets of comparative experiments using the MusicTransformer \cite{Huang2018} and MelodyT5 \cite{Wu2024} models as baselines. In each experiment, we randomly generate approximately 50 songs for each model and calculated objective metrics, which are displayed in the table. When evaluating the quality of generated music, we consider values that are closer to real data as better. As shown in the \cref{tab:pitch-entropy}, our model's generated music is closer to the real values in terms of pitch entropy, outperforming the other models. Our model also excels in style consistency and rhythm consistency. Notably, in terms of mode consistency, over 70\% of the music generated by our model exhibits a detectable modal structure, and more than 60\% of the music performs well in mode consistency. The above metrics are shown in the \cref{tab:performance}.

\subsection{Subjective Listening Test}
\label{sec:subjective}
To evaluate the quality of the music samples generated by the model, we design a subjective listening test. We recruit 10 music enthusiasts from social networks, each of whom plays at least one musical instrument. Each participant is asked to listen to 10 generated audio samples. They rate the samples based on three criteria: coherence, richness, and style, with scores ranging from 0 to 10. In the subjective evaluation results, MusicMamba outperforms all baseline models in coherence, richness, and style, showing the best overall performance.

\section{Conclusion}

In this paper, we proposed a novel architecture that combines the long-range dependency modeling capability of the Mamba Block with the global structure-capturing ability of the Transformer Block. We also designed the Bidirectional Mamba Fusion Layer to effectively integrate local and global information. By introducing the REMI-M representation, we were able to capture and generate modal features in Chinese traditional music with greater accuracy. Experimental results demonstrate that the combination of REMI-M and MusicMamba more precisely reproduces and generates specific modes in Chinese traditional music. The generated music outperforms traditional baseline models in terms of stylistic consistency and quality. Our research provides a new direction and technical foundation for exploring more complex modes in various ethnic music genres and generating melodies with distinctive styles through the incorporation of traditional instruments.


\newpage
\onecolumn
\twocolumn

\bibliographystyle{IEEEbib}
\bibliography{ref}

\end{document}